\documentclass[twocolumn]{aastex63}
\received{January 1, 2018}
\revised{January 7, 2018}
\accepted{\today}

\shorttitle{Evidence for X-ray resonance scattering in SNR N49}
\shortauthors{Amano et al.}

\begin{document}

\title{Evidence for Resonance Scattering in the X-ray Grating Spectrum of the Supernova Remnant N49}

\correspondingauthor{Yuki Amano}
\email{amano.yuki.t76@kyoto-u.jp}

\author{Yuki Amano}
\affil{Department of Physics, Kyoto University, Kitashirakawa Oiwake-cho, Sakyo, Kyoto, Kyoto 606-8502, Japan}

\author{Hiroyuki Uchida}
\affiliation{Department of Physics, Kyoto University, Kitashirakawa Oiwake-cho, Sakyo, Kyoto, Kyoto 606-8502, Japan}

\author{Takaaki Tanaka}
\affiliation{Department of Physics, Kyoto University, Kitashirakawa Oiwake-cho, Sakyo, Kyoto, Kyoto 606-8502, Japan}

\author{Liyi Gu}
\affiliation{RIKEN High Energy Astrophysics Laboratory, 2-1 Hirosawa, Wako, Saitama 351-0198, Japan}
\affiliation{SRON Netherlands Institute for Space Research, Sorbonnelaan 2, 3584 CA Utrecht, the Netherlands}

\author{Takeshi Go Tsuru}
\affiliation{Department of Physics, Kyoto University, Kitashirakawa Oiwake-cho, Sakyo, Kyoto, Kyoto 606-8502, Japan}



\begin{abstract}
Resonance scattering (RS) is an important process in astronomical objects, because it affects measurements of elemental abundances and distorts surface brightness of the object.
It is predicted that RS can occur in plasmas of supernova remnants (SNRs).
Although several authors reported hints of RS in SNRs, no strong observational evidence has been established so far.
We perform a high-resolution X-ray spectroscopy of the SNR N49 with the Reflection Grating Spectrometer aboard XMM-Newton.
The RGS spectrum of N49 shows a high G-ratio of \ion{O}{7} He$\alpha$ lines as well as \ion{O}{8} Ly$\beta$/$\alpha$ and \ion{Fe}{17} (3s--2p)/(3d--2p) ratios 
which cannot be explained by the emission from a thin thermal plasma.
These line ratios can be well explained by the effect of RS.
Our result implies that RS has a large impact particularly on a measurement of the oxygen abundance.
\end{abstract}



\section{Introduction} \label{sec:intro}
X-ray imaging spectroscopy of astronomical objects provides us with many insights into their chemical evolution and formation mechanism. 
This is partially because X-ray emitting plasmas are often optically thin, which allows us to directly estimate its elemental abundances and their spatial distributions.
However, resonance lines such as \ion{O}{7} He$\alpha$ and \ion{Fe}{17} L$\alpha$ may suffer from effects of scattering (resonance scattering: RS), as discussed in the cases of galaxies \citep[e.g.,][]{Xu2002}, solar active regions \citep[e.g.,][]{Rugge1985}, and galaxy clusters \citep[e.g.,][]{Hitomi2018}.

RS is an apparent scattering phenomenon due to an absorption and re-emission of line photons by ions.
Since the RS effect apparently reduces intensities of some lines 
and/or distorts profiles of surface brightness \citep{Shigeyama1998}, ignoring its contribution can sometimes lead to, for example, biases in elemental abundance measurements.
On the other hand, if RS is significant, quantifying its contribution will allow us to measure several important parameters such as micro-turbulence velocities \citep{dePlaa2012} and 
absolute abundances \citep{Waljeski1994}.

\cite{Kaastra1995} predicted that RS of X-ray photons can occur in a plasma with a large depth along the line of sight such as a rim of supernova remnants (SNRs).
Several observational signatures of RS have been reported such as a high forbidden-to-resonance ratio of \ion{O}{7} He$\alpha$ obtained from grating spectra of  
DEM~L71 \citep{vanderHeyden2003} and N23 \citep{Broersen2011}. 
A difference in surface brightness between forbidden and resonance lines also supports the presence of RS \citep[e.g.,][]{vanderHeyden2003}. 
Based on a Suzaku observation of the Cygnus Loop, \cite{Miyata2008} claimed that a depleted abundance of O may be partially explained by RS.
A recent grating observation of the Loop also hints at a possibility of RS \citep{Uchida2019}.
These studies suggest that the RS effect may potentially be significant in SNRs.
However, no strong observational evidence has been established so far. 

N49 is a middle-aged \citep[$\sim4800$~yr;][]{Park2012} SNR located in the Large Magellanic Cloud (LMC).
Although the origin of N49 is somewhat controversial, most of recent researches support that it originated from a core-collapse explosion based on 
a possible association with the soft gamma-ray repeater (SGR) 0526$-$66 \citep{Cline1982} and the presence of a dense interstellar medium (ISM) \citep[e.g.,][]{Banas1997, Yamaguchi2014}. 
The thermal X-ray emission is explained by a mixture of two components; metal-rich ejecta and a shock-heated ISM \citep[][]{Park2003, Park2012, Uchida2015}.
It is also notable that N49 is in an overionized state \citep{Uchida2015} and is interacting with dense molecular clouds on the eastern side \citep[][]{Banas1997, Yamane2018}.
Due to the interaction with molecular clouds, the thermal X-ray emission of N49 is particularly bright on the southeastern rim.
N49 is an attractive object since \cite{Kaastra1995} pointed out that the effect of RS can show up if an SNR has such a spherically asymmetric structure.

Here, we present a high-resolution X-ray spectroscopy of N49 with the Reflection Grating Spectrometer (RGS) aboard XMM-Newton.
The obtained G-ratio of \ion{O}{7} He$\alpha$ and other line ratios as well imply a non-negligible contribution of RS in N49.
Throughout the paper, errors are given at a 68\% confidence level.
We assume the distance to N49 (LMC) to be 50~kpc \citep{Pietrzy2013}.

\section{Observation and Data Reduction} \label{sec:obs}

\begin{figure}
\begin{center}
 \includegraphics[width=7.0cm]{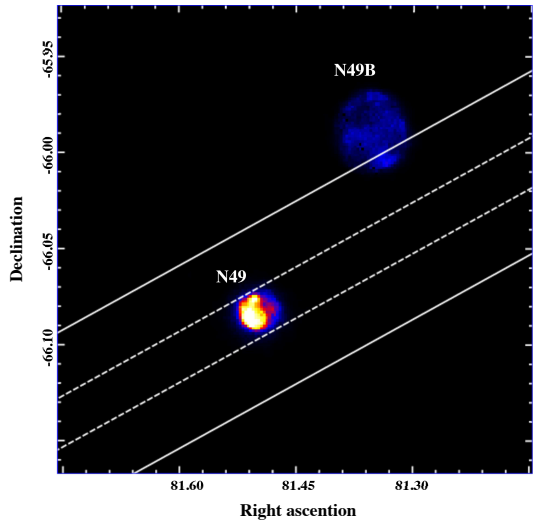} 
\end{center}
\caption{EPIC-MOS image (0.4--8.0~keV) of N49. The cross-dispersion width of the RGS (5~arcmin) is in between the white solid line. The spectral extraction region is enclosed by the white dashed line. N49B is another SNR near N49.}
\label{fig:mosimage}
\end{figure}

N49 was observed with the XMM-Newton satellite \citep{Jansen2001} in 2001 (Obs.ID 0113000201) and 2007 (Obs.ID 0505310101).
The observation in 2007 was performed with a roll angle which placed N49 and a nearby SNR, N49B, 
along the dispersion direction of the RGS, making it difficult to extract RGS spectra of N49. 
We thus analyzed only the data obtained in 2001.
For our spectral analysis, we used the RGS \citep{denHerder2001} and the European Photon Imaging Camera MOS \citep{Turner2001} data. 
We reduced the data using XMM Science Analysis Software version 16.1.0.
The RGS data were processed with the RGS pipeline tool {\tt rgsproc}.
To discard periods of background flares, we applied good time intervals based on the count rate in CCD9, which is the closest to the optical axis of the mirror 
and has the least X-ray counts from the source. 
and most affected by the background flares.
The resulting effective exposure time is 11~ks for both RGS1 and RGS2. 
Since the second order spectra are of low statistical quality, we analyzed only the first order spectra.

\begin{figure*}
\begin{center}
 \includegraphics[width=15.0cm]{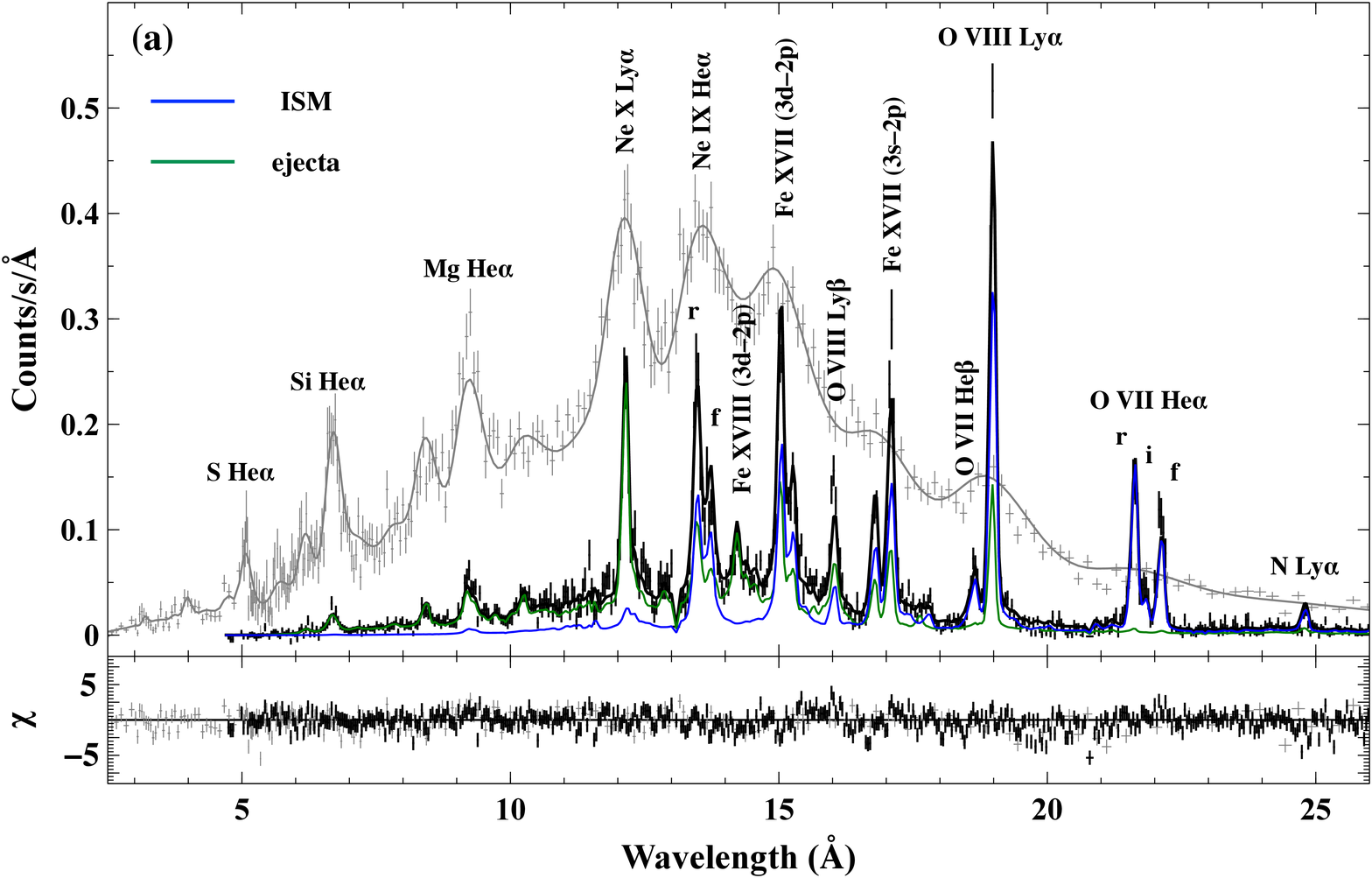} 
\end{center}

\begin{center}
 \includegraphics[width=15.0cm]{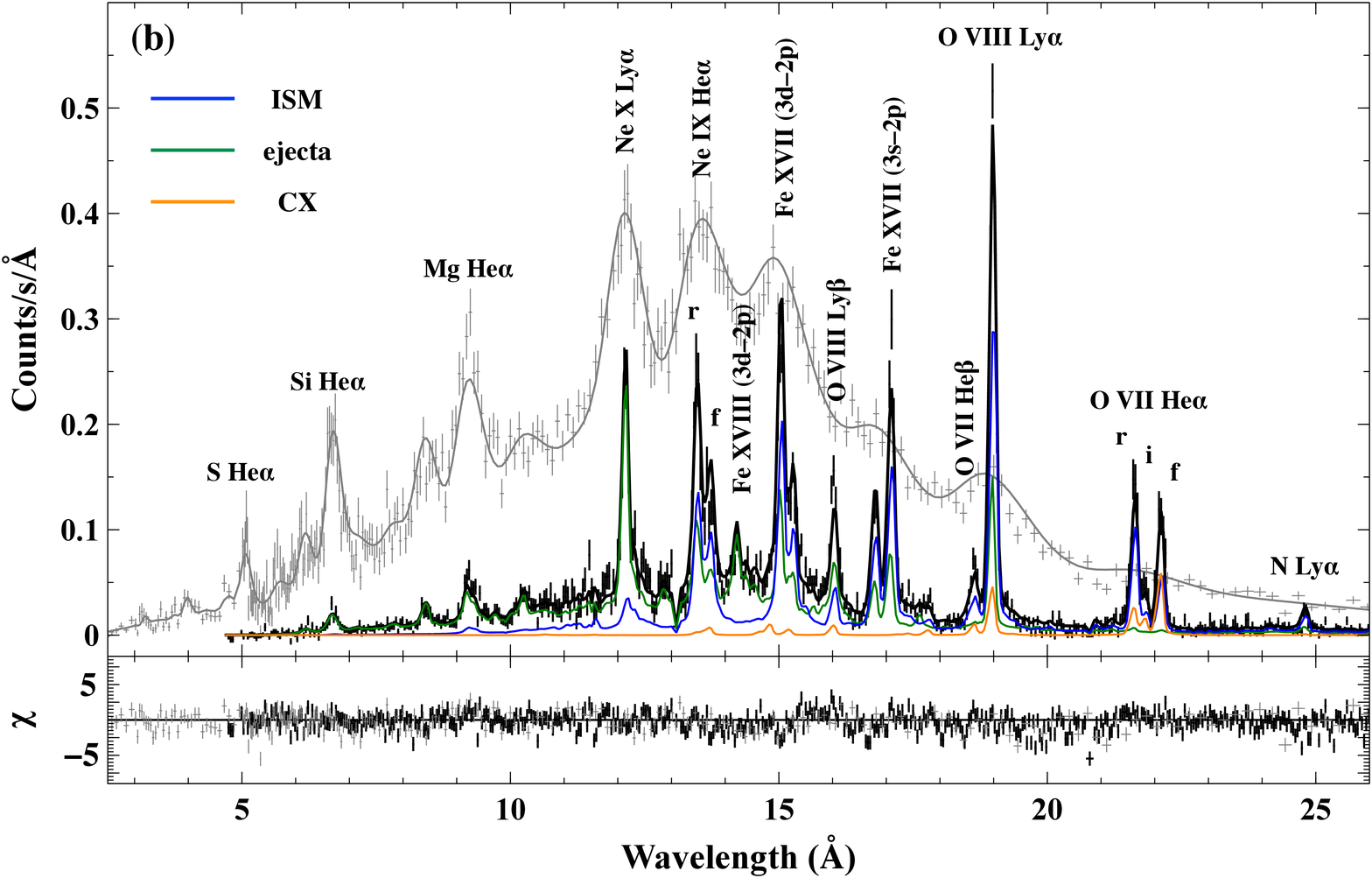} 
\end{center}
\caption{ 
RGS1$+$2 (black) and MOS1 (gray) spectra of N49.  Overlaid in panels (a) and (b) are the best-fit ``NEI'' and ``NEI$+$CX'' models, respectively.
}
\label{fig:spec}
\end{figure*}

\begin{deluxetable*}{llccc}
\tablecaption{Best-fit parameters of the N49 spectrum \label{tab:parameters}}
\tablecolumns{5}
\tablenum{1}
\tablehead{
\colhead{Component} &
\colhead{Parameters (unit)} &
\colhead{NEI} &
\colhead{NEI + CX} &
\colhead{NEI - Gaussians (RS)}
}
\startdata
Absorption & ${N_{\rm H}({\rm MW})~(10^{22}~{\rm cm^{-2}})}$ & $0.6$ (fixed)& $0.6$ (fixed) & $0.6$ (fixed) \\
 & ${N_{\rm H}({\rm LMC})~(10^{22}~{\rm cm^{-2}})}$  &  $3.1 \pm 0.1$ & $2.8 \pm 0.1$ & $3.3 \pm 0.1$ \\
ISM & ${kT_{\rm e}}~({\rm keV})$ & $0.204 \pm 0.003$ & $0.230 \pm 0.006$ & $0.205 \pm 0.003$ \\
 & ${n_{\rm e}t~(10^{11}~{\rm cm^{-3}~s}}$) & $>10$ & $>10$ & $>10$\\
 & ${EM~(10^{56}~{\rm cm^{-3}})}$ & $260 \pm 20$ & $170 ^{+20}_{-10}$ & $322^{+23}_{-22}$ \\
 Ejecta & ${kT_{\rm e}~({\rm keV})}$ & $0.61 \pm 0.01$ & $0.63 \pm 0.01$ & $0.56^{+0.08}_{-0.01}$ \\
  & ${kT_{\rm init}~({\rm keV})}$ & $11$~(fixed) & $11$~(fixed) & $11$~(fixed) \\
  & ${n_{\rm e}t~(10^{11}~{\rm cm^{-3}~s}}$) & $7.2^{+0.5}_{-0.3}$ & $7.2^{+0.4}_{-0.3}$ & $6.9^{+1.0}_{-0.4}$ \\
  & O (= C = N) & $0.71^{+0.16}_{0.06}$ & $0.67^{+0.20}_{-0.14}$ & $1.3 \pm 0.1$ \\
  & Ne & $0.96^{+0.09}_{-0.07}$ & $0.94^{+0.09}_{-0.06}$ & $1.0 \pm 0,1$ \\
  & Mg & $0.75 \pm 0.07$ & $0.72^{+0.07}_{-0.06}$ & $0.75 {\rm (fixed)}$ \\
  & Si & $0.87^{+0.08}_{-0.06}$ & $0.85^{+0.08}_{-0.06}$ & $0.87 {\rm (fixed)}$ \\
  & S & $1.2 \pm 0.1$ & $1.2 \pm 0.1$ & $1.2 {\rm (fixed)}$ \\
  & Ar & $1.8 \pm 0.5$ & $1.8^{+0.4}_{-0.3}$ & $1.8 {\rm (fixed)}$ \\
  & Fe & $0.32 \pm 0.03$ & $0.29^{+0.03}_{-0.02}$ & $0.32 \pm 0.01$ \\
  & ${EM~(10^{56}~{\rm cm^{-3}})}$ & $58 \pm 4$ & $57^{+5}_{-2}$ & $53 \pm 0.1$ \\
CX & ${kT_{\rm e}~({\rm keV})}$ & \nodata & (= value of the ISM component) & \nodata \\
  & abundance & \nodata & (= abundances of the ISM component) & \nodata \\
  & ${v_{\rm collision}}~({\rm km}~{\rm s}^{-1})$ & \nodata & $270 \pm 110$ &  \nodata\\
  & ${EM~(10^{56}~{\rm cm^{-3}})}$ & \nodata & $47^{+25}_{-13}$ & \nodata \\ 
  Gaussian: \ion{Ne}{9} He$\alpha$ & Normalization ($10^{44}~{\rm ph~s^{-1}}$) & \nodata & \nodata & $0.36 \pm 0.09$ \\ 
  ~~~~~~~~~~~~~~\ion{Fe}{17} L$\alpha$ ($\sim 15.0~{\rm \AA}$) &  & \nodata & \nodata & $0.92 \pm 0.11$ \\
  ~~~~~~~~~~~~~~\ion{Fe}{17} L$\alpha$ ($\sim 15.3~{\rm \AA}$) &  & \nodata & \nodata & $< 6.9 \times10^{-2}$ \\
  ~~~~~~~~~~~~~~\ion{O}{8} Ly$\beta$ &  & \nodata & \nodata & $< 1.3 \times10^{-2}$ \\
  ~~~~~~~~~~~~~~\ion{Fe}{17} L$\alpha$ ($\sim 16.8~{\rm \AA}$) &  & \nodata & \nodata & $0.50 \pm 0.12$ \\
  ~~~~~~~~~~~~~~\ion{Fe}{17} L$\alpha$ ($\sim 17.1~{\rm \AA}$) &  & \nodata & \nodata &  $< 1.5 \times10^{-3}$  \\
  ~~~~~~~~~~~~~~\ion{O}{7} He$\beta$ &  & \nodata & \nodata &  $< 5.8 \times10^{-2}$ \\
  ~~~~~~~~~~~~~~\ion{O}{8} Ly$\alpha$ &  & \nodata & \nodata & $4.6 \pm 0.3$ \\
  ~~~~~~~~~~~~~~\ion{O}{7} He$\alpha$(r) &  & \nodata & \nodata & $2.5 \pm 0.5$ \\ 
  ~~~~~~~~~~~~~~\ion{O}{7} He$\alpha$(i) &  & \nodata & \nodata & $< 3.0 \times10^{-3}$  \\ 
 \hline
  & C-statistic/d.o.f. & $4700/3211$ & $4660/3214$ & $4611/3190$ \\
  \enddata
  \tablecomments{The elemental abundances are given with respect to the solar values by \cite{Lodders2009}.}
   \tablecomments{See Table~\ref{tab:transmission} for the line centroid wavelengths of the Gaussians.}
  \end{deluxetable*}

\section{Analysis} \label{sec:analysis}
We analyzed the spectra using version 3.04.0 of the SRON SPEX package \citep{Kaastra1996} with the maximum likelihood C-statistic \citep{Cash1979,Kaastra2017}.  
The RGS spectra were fitted simultaneously with those of MOS1 and 2.
To account for the spatial broadening of the source, we multiplied spectral models with the SPEX model {\tt lpro}, to which we input the MOS1 image of the source. 
Our models have two absorption models: one for the Milky Way and the other for the LMC. 
The column density of the former was fixed to $6 \times 10^{20}~{\rm cm^{-2}}$ \citep{DickeyLockman1990} whereas that of the latter is left free.
The elemental abundances for the LMC absorption were fixed to values found in literature \citep[$\sim$~0.3 solar;][]{Russell1992, Schenck2016}.

Figure~\ref{fig:spec} shows MOS1 and combined RGS1$+$2 spectra of N49. 
Prominent lines are detected at $\sim$12~\AA \ (\ion{Ne}{10} Ly$\alpha$), $\sim$13.5~\AA \ (\ion{Ne}{9} He$\alpha$),  $\sim$15~\AA \ (\ion{Fe}{17} L$\alpha$; 3d--2p), $\sim$16~\AA \ (\ion{O}{8} Ly$\beta$), $\sim$17~\AA \ (\ion{Fe}{17} L$\alpha$; 3s--2p),  $\sim$20~\AA \ (\ion{O}{7} He$\beta$, \ion{O}{8} Ly$\alpha$), $\sim$22~\AA \ (\ion{O}{7} He$\alpha$).
We applied a two-component nonequilibrium ionization (NEI) model \citep[{\tt neij};][]{KaastraJansen1993} absorbed by neutral gas to the spectra. 
The NEI model consists of emissions from the overionized hot ejecta and an ionizing cool component originating from the swept-up ISM \citep{Uchida2015}.  
Although SGR~0526$-$66 cannot be spatially resolved with the RGS, according to the result of \cite{Uchida2015}, the emission is negligible compared to that from N49 in the energy band covered by the RGS (0.3--2.0~keV).

Free parameters of the NEI components include the electron temperature ($kT_{\rm e}$), ionization time scale ($n_{\rm e}t$, where $n_{\rm e}$ and $t$ are the electron number density and the elapsed time since shock heating or rapid cooling, respectively), and emission measure ($n_{\rm e}n_{\rm H}V$).
In addition to these parameters, the abundances of O (=N=C), Ne, Mg, Si, S, Ar, and Fe (=Ni) of the ejecta were set free. 
The {\tt neij} model has another parameter $kT_{\rm init}$, which represents an initial temperature before a rapid cooling or a shock heating of the plasma.
\cite{Uchida2015} determined $kT_{\rm init}$ for the ejecta component mainly from the ionization state of the Fe K$\alpha$ emission at $\sim$6.6~keV.
Since the line centroid is out of the wavelength band of the RGS data, we fixed $kT_{\rm init}$ of the ejecta component at 11~keV based on \cite{Uchida2015}.
On the other hand, the $kT_{\rm init}$ of the ISM component is fixed at $\sim 0 ~{\rm keV}$.

\begin{figure}
\begin{center}
 \includegraphics[width=8.0cm]{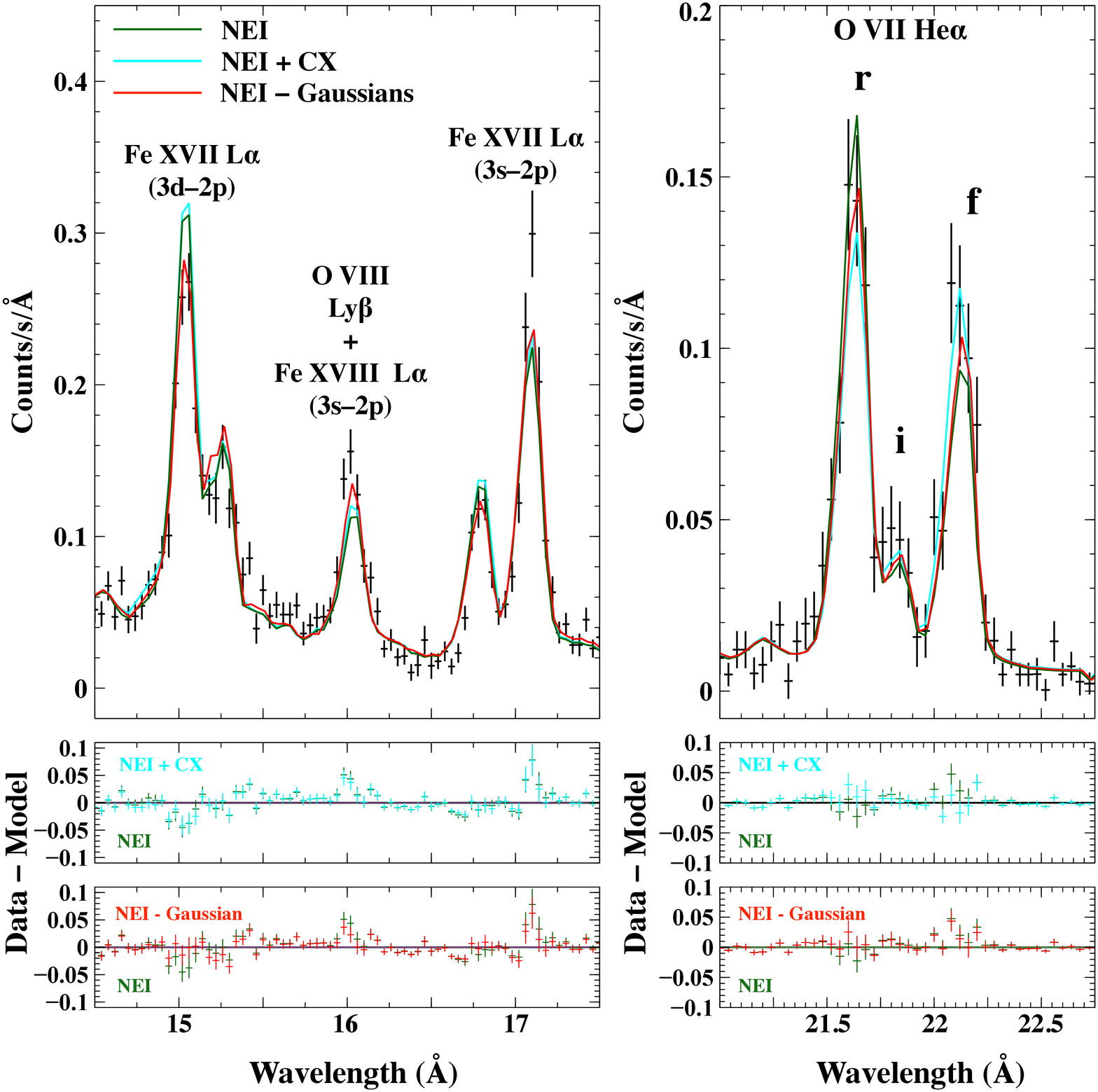} 
\end{center}
\caption{
Close-up views of the RGS spectrum (Figure~\ref{fig:spec}) around \ion{Fe}{17} L$\alpha$ and \ion{O}{8} Ly$\beta$ (left) 
and  \ion{O}{7} He$\alpha$ (right). 
The dark green, cyan, and red solid lines in the top panels represent the best-fit ``NEI'', ``NEI$+$CX'', and ``NEI$-$Gaussians'' models, respectively.
The middle and bottom panels show residuals from the models with the same color scheme as the top panels.
}
\label{fig:FeOspec}
\end{figure}

Figure~\ref{fig:spec} (a) shows the result of the spectral fitting with the two-component NEI model (hereafter, NEI model).
The best-fit parameters are listed in Table~\ref{tab:parameters}.
The model well reproduces the overall MOS spectrum; the best-fit parameters of $kT_{\rm e}$ and $n_{\rm e}t$ of the ejecta component are consistent with those obtained by \cite{Uchida2015}.
Focusing on the RGS spectrum in Figure~\ref{fig:FeOspec}, however, we found discrepancies between the model and the data especially around the \ion{O}{7} triplet, \ion{O}{8} Ly$\beta$, and \ion{Fe}{17} L$\alpha$ series.

\begin{figure}
\begin{center}
 \includegraphics[width=7.0cm]{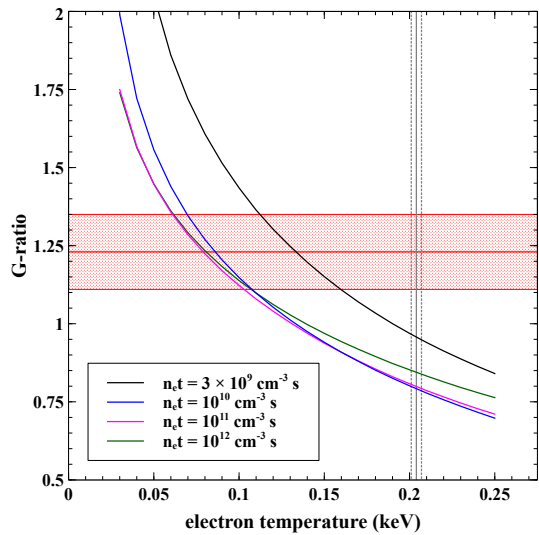} 
\end{center}
\caption{
Relation between $kT_{\rm e}$ and the G-ratio of \ion{O}{7} He$\alpha$.
The solid curves represent the G-ratio expected for an ionizing plasma emission predicted by the {\tt neij} model in SPEX. 
The colors of each line indicate $n_{\rm e}t$ assumed. 
The red hatched area denotes the G-ratio derived from the observed line ratio and its statistical error. 
The gray lines indicates $kT_{\rm e}$ from the best-fit NEI model. 
}
\label{fig:gratio}
\end{figure}

We first focus on the \ion{O}{7} He$\alpha$ line, more specifically on its G-ratio ($F+I)/R$, where $F$, $I$, and $R$ are intensities of 
the forbidden, intercombination, and resonance lines, respectively.
The G-ratio of the \ion{O}{7} He$\alpha$ line strongly depends on $kT_{\rm e}$ as presented in Figure~\ref{fig:gratio}. 
We assume here that the emitting plasma is in an ionizing state since most of the \ion{O}{7} He$\alpha$ line emission of N49 is attributed to 
the shocked ISM (Figure~\ref{fig:spec} (a)). 
Fitting with four Gaussians, we estimated the G-ratio of the \ion{O}{7} He$\alpha$ line observed in N49, which is plotted in Figure~\ref{fig:gratio}. 
The observed G-ratio obviously requires an unreasonably low $kT_{\rm e}$  ($< 0.15~{\rm keV}$) with any reasonable range of $n_{\rm e}t$, indicating 
necessity of some physical process to enhance the G-ratio.

\begin{figure*}
\begin{center}
 \includegraphics[width=15.0cm]{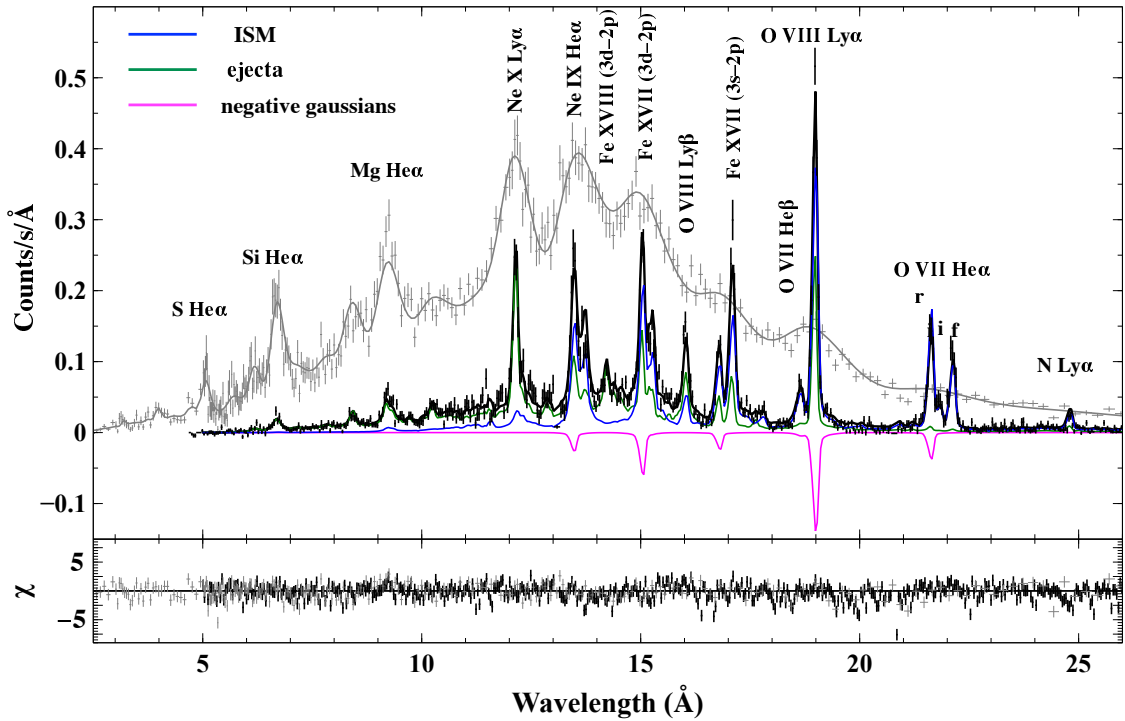} 
\end{center}
\caption{
Same as Figure~\ref{fig:spec} but with the ``${\rm NEI}-{\rm Gaussians}$'' model. 
The magenta solid line represents the negative Gaussians.}
\label{fig:rsfit}
\end{figure*}

Charge exchange (CX) is one of the possible processes to make the G-ratio higher. 
Recent X-ray spectroscopy studies with gratings found enhanced $F/R$ ratios of \ion{O}{7} in Puppis~A \citep{Katsuda2012}, and the Cygnus Loop \citep{Uchida2019}.
They claimed that the anomalous line ratios are explained by CX X-ray emission.
As suggested by \cite{Uchida2019}, the CX emission can be enhanced particularly in a region where a shock is interacting with dense gas.
CX is therefore promising in the case of N49 since it is interacting with surrounding molecular clouds \citep{Yamane2018}.
We thus added a CX model \citep{Gu2016a} to the NEI model.
All the abundances and $kT_{\rm e}$ for the CX component were coupled to those of the ISM component.
The collision velocity $v_{\rm collision}$ ({\it i.e.} shock velocity) was allowed to vary.
We assumed a multiple collision case in which an ion continuously undergoes CX until it becomes neutral.  
The best-fit ``NEI$+$CX'' model is displayed in Figure~\ref{fig:spec} (b), and the parameters are listed in Table~\ref{tab:parameters}.
Although the addition of the CX component improves the residuals at the \ion{O}{7} triplet, the discrepancies still remain around \ion{O}{8} Ly$\beta$ and \ion{Fe}{17} L$\alpha$ series (Figure~\ref{fig:FeOspec}).
Therefore, we conclude that the NEI + CX model is insufficient to describe the spectrum of N49.

RS is another possible process that would be responsible for the enhanced G-ratio. 
If photons of the resonance line are scattered off the line of sight due to RS, observed $R$ would become lower and the G-ratio would be enhanced.  
The observed high \ion{Fe}{17} (3s--2p)/(3d--2p) and \ion{O}{8} Ly$\beta$/$\alpha$ ratios can also be caused by RS, as pointed out by previous studies \citep[e.g.,][]{Xu2002, Hitomi2018}.
To quantify the contribution of the scattering effect, we added negative Gaussians at wavelengths where lines of the ISM component are prominent in the best-fit NEI model:  the \ion{O}{7} He$\alpha$ intercombination, \ion{O}{7} He$\alpha$ resonance, \ion{O}{7} He$\beta$, \ion{O}{8} Ly$\alpha$, \ion{O}{8} Ly$\beta$, \ion{Fe}{17} L$\alpha$, 
and \ion{Ne}{9} He$\alpha$ resonance lines.
Since the oscillator strengths of the \ion{O}{7} and \ion{Ne}{9} forbidden lines are several orders of magnitude smaller than those of the other lines, we assume that the scattering effect  is negligible for the forbidden lines.
The free parameters of the NEI components are the electron temperature ($kT_{\rm e}$), ionization time scale ($n_{\rm e}t$), emission measure ($n_{\rm e}n_{\rm H}V$).
The abundances of O (=N=C), Ne, and Fe (=Ni) of the ejecta were set free and the abundances of Mg, Si, S, and Ar were fixed to the values obtained in the NEI model fit. 
The best-fit  ``${\rm NEI}-{\rm Gaussians}$'' model is displayed in Figure \ref{fig:rsfit} and the best-fit parameters are listed in Table \ref{tab:parameters}.  

\begin{deluxetable*}{lccc}
\tablecaption{Transitions, centroid wavelengths, and transmission factors of each line.\label{tab:transmission}}
\tablecolumns{4}
\tablenum{2}
\tablewidth{0pt}
\tablehead{
\colhead{Line} &
\colhead{Transition} &
\colhead{Line Centroid (${\rm \AA}$)\tablenotemark{a}} &
\colhead{Transmission Factor\tablenotemark{b}} 
}
\startdata
   \ion{Ne}{9} He$\alpha$ &$1{\rm s}2{\rm p}\, ^1{\rm P}_1 \textrm{--}  1{\rm s}^2 \, ^1{\rm S}_0 $&13.45& $0.79 \pm 0.11$ \\
   \ion{Fe}{17} L$\alpha$ &$ 2{\rm s}^22{\rm p}^53{\rm d}\, ^1{\rm P}_1 \textrm{--} 2{\rm s}^22{\rm p}^6\,^1{\rm S}_0 $&15.02& $0.70^{+0.04}_{-0.08}$ \\
   \ion{Fe}{17} L$\alpha$ &$ 2{\rm s}^22{\rm p}^53{\rm d}\,^3{\rm D}_1 \textrm{--} 2{\rm s}^22{\rm p}^6\,^1{\rm S}_0  $&15.26& $>0.85$ \\
   \ion{O}{8} Ly$\beta$ &$ 3{\rm p}\,^2{\rm P}_{3/2}  \textrm{--} 1{\rm s}\,^2{\rm S}_{1/2} $&16.01& $>0.82$ \\
   \ion{Fe}{17} L$\alpha$ &$ 2{\rm s}^22{\rm p}^53{\rm s}\,^3{\rm P}_1 \textrm{--} 2{\rm s}^22{\rm p}^6\,^1{\rm S}_0  $&16.78& $0.73^{+0.11}_{-0.18}$ \\
   \ion{Fe}{17} L$\alpha$ &$ 2{\rm s}^22{\rm p}^53{\rm s}\,^1{\rm P}_1 \textrm{--} 2{\rm s}^22{\rm p}^6\,^1{\rm S}_0  $&17.05& $>0.90$ \\
   \ion{O}{7} He$\beta$ &$ 1{\rm s}3{\rm p}\,^1{\rm P}_1 \textrm{--} 1{\rm s}^2\,^1{\rm S}_0 $&18.63& $>0.79$ \\
   \ion{O}{8} Ly$\alpha$ &$2{\rm p}\,^2{\rm P}_{3/2} \textrm{--} 1{\rm s}\,^2{\rm S}_{1/2}$&18.97& $0.46^{+0.02}_{-0.03}$ \\
   \ion{O}{7} He$\alpha$(r) &$1{\rm s}2{\rm p}\,^1{\rm P}_1 \textrm{--} 1{\rm s}^2\,^1{\rm S}_0$&21.60& $0.79 \pm 0.08$ \\ 
   \ion{O}{7} He$\alpha$(i) &$1{\rm s}2{\rm p}\,^3{\rm P}_1 \textrm{--} 1{\rm s}^2\,^1{\rm S}_0$&21.81& $>0.76$  \\
\enddata
 \tablenotetext{a}{Taken from SPEX.}
 \tablenotetext{b}{Calculated from the best-fit parameters of the ``${\rm NEI}-{\rm Gaussians}$'' model.}
\end{deluxetable*}

\section{Discussion} \label{sec:results}
We compare transmission factors estimated based on the result from the ``${\rm NEI}-{\rm Gaussians}$'' model fit 
with those expected for RS. 
The transmission factor $p$ is defined as
\begin{equation}
\label{eq:ep}
p = \frac{A - \Delta A}{A},
\end{equation}
where $A$ is the total number of photons emitted by the plasma, and $\Delta A$ is the number of photons scattered out of our line of sight.
We can derive $p$ for each line from the data, given that the best-fit normalizations of the negative Gaussians and the line intensities of the ISM component 
correspond to $\Delta A$ and $A$, respectively. 
The transmission factors derived from the data are listed in Table \ref{tab:transmission}.

Referring to \cite{Kaastra1995}, we calculate the transmission factors  in a case where RS effectively occurs. 
Under the slab approximation as a simple geometrical model of the SNR rim, \cite{Kaastra1995} used a single-scattering treatment where a photon completely escapes from the line of sight at every scattering event.
As shown by \cite{Park2003}, the ISM plasma of N49 has a particularly bright emission at the southeastern rim.
Assuming that the RS occurs dominantly at the southeastern rim, we adopt the same assumption as \cite{Kaastra1995}.
Then, the transmission factor is written as 
\begin{equation}
\label{eq:ep2}
p = \frac{1}{1 + 0.43 \tau},
\end{equation}
where $\tau$ is the optical depth of the ISM plasma \citep{Kastner1990}.
The optical depth $\tau$ at the line centroid is given by \cite{Kaastra1995} as
\begin{equation}
\label{eq:RS_tau}
 \tau = \frac{4.24 \times 10^{26} f N_{\rm H} \left(\frac{n_{\rm i}} {n_{\rm z}}\right) \left(\frac{n_{\rm z}}{n_{\rm H}}\right) \left(\frac{M}{T_{\rm keV}}\right)^{1/2}}{E_{\rm eV}\left(1 + \frac{0.0522 M v_{100}^2}{T_{\rm keV}}\right)^{1/2}},
\end{equation}
where $f$ is the oscillator strength of the line, $E_{\rm eV}$ is the line centroid energy in eV, $N_{\rm H}$ is the hydrogen column density in ${\rm cm^{-2}}$, $n_{\rm i}$ is the number density of the ion, $n_{\rm Z}$ is the number density of the element, $M$ is the atomic weight of the ion, $T_{\rm keV}$ is the ion temperature in keV, and $v_{100}$ is the micro-turbulence velocity in units of $100~{\rm km}~{\rm s}^{-1}$.
We assumed a thermal equilibrium between all ions and electrons and neglected the micro-turbulence velocity.
The oscillator strengths and ion fractions for each element were taken from SPEX.
In our case, the absorption column density ($N_{\rm H}$) is the only free parameter.

\begin{figure*}
\begin{center}
 \includegraphics[width=15.0cm]{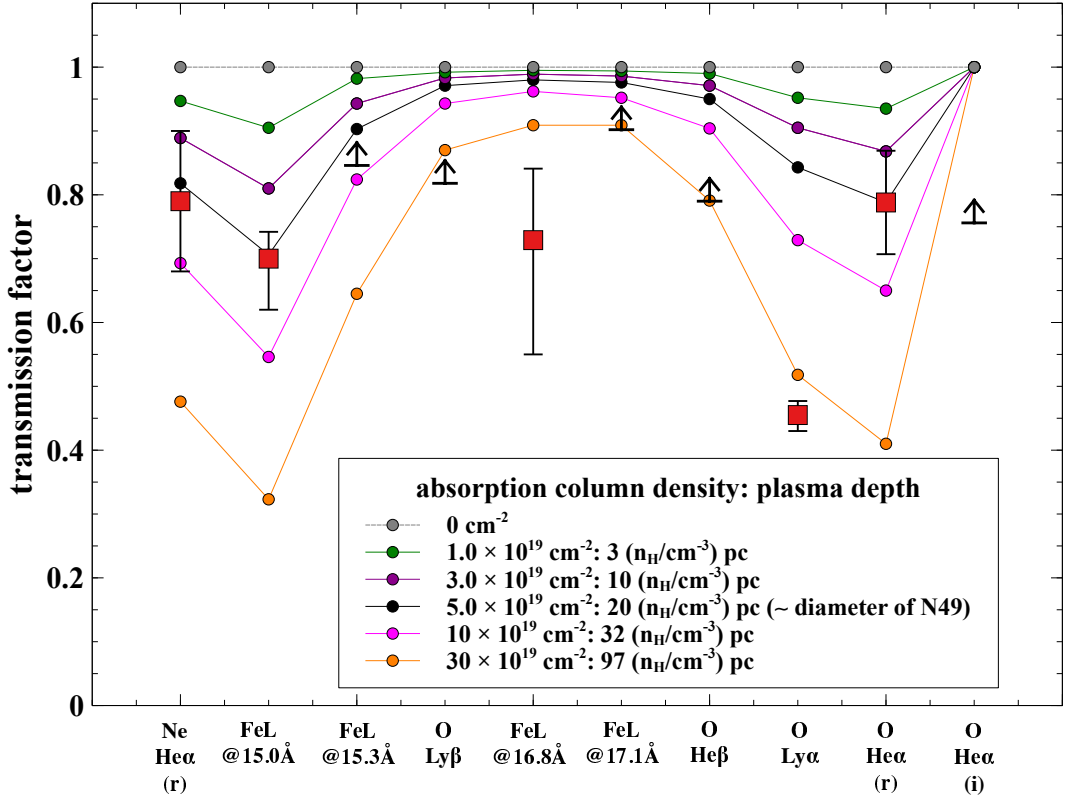} 
\end{center}
\caption{
Transmission factors for each line.
The squares and arrows are the values and lower limits derived from the observational data 
whereas the circles are those expected for RS.
The colors indicate $N_{\rm H}$ assumed in the calculation.
}
\label{fig:epplot}
\end{figure*}

In Figure~\ref{fig:epplot}, we compare the transmission factors estimated from the data and those calculated with equation~(\ref{eq:ep2}). 
They are roughly consistent if $N_{\rm H}$ is (3.0--10)$\times10^{19}~{\rm cm^{-2}}$, which corresponds to a plasma depth of (10--34)$\times(n_{\rm H}/{\rm cm^{-3}})$~pc.
Since the plasma depth is comparable to the diameter of N49, $\sim20$~pc, the result supports that RS occurs at the rim of N49. 
We note that we would underestimate the O abundance by about a factor of 1.8 if we do not take into account the RS effect (Table \ref{tab:parameters}). 
This demonstrates the importance of RS in measuring elemental abundances as already pointed out by, e.g., \cite{Kaastra1995} and \cite{Miyata2008}. 

In Figure~\ref{fig:epplot}, we found that  \ion{O}{8} Ly$\alpha$ requires a higher column density than the other lines.
A possible explanation would be that RS occurs also in the Galactic Halo (GH), as proposed by \cite{Gu2016b}.  
According to \cite{Nakashima2018}, GH spectra are represented by a collisional ionization equilibrium (CIE) plasma model with $kT_{\rm e} \sim 0.26$~keV. 
Since such a plasma has a larger optical depth for the \ion{O}{8} Ly$\alpha$ line than that of the \ion{O}{7} He$\alpha$ resonance line, the GH should selectively reduce the intensity of \ion{O}{8} Ly$\alpha$.
We applied the CIE absorption model, {\tt hot} \citep{dePlaa2004,Steenbrugge2005}, to the ${\rm NEI}-{\rm Gaussians}$ model.
The electron temperature and Fe abundance of the {\tt hot} model are fixed to 0.26~keV and 0.56 solar, respectively (the other elemental abundances are fixed to the solar values), by referring to \cite{Nakashima2018}.
We found that the discrepancy cannot be reduced with the model, where the transmission factors of \ion{O}{8} Ly$\alpha$ and \ion{O}{7} He$\alpha$ 
are calculated to be $\sim0.43$ and $\sim0.80$, respectively.

The cause of the observed high \ion{O}{8} Ly$\beta$/$\alpha$ ratio (the measured value is 0.18 as an upper limit) is not clear.
Uncertainties in the model of the Fe-L lines \citep[e.g.,][]{Gu2019} might partially explain the result since \ion{O}{8} Ly$\beta$ overlaps with the \ion{Fe}{18} L$\alpha$ line.
Another possibility to explain the high \ion{O}{8} Ly$\beta$/$\alpha$ ratio is an effect related to RS.
As discussed by \cite{Chevalier1980}, Ly$\beta$ can be converted to H$\alpha$ by a $3p$-$2s$ transition in a collisionless shock through RS.
However, this is less plausible because this effect reduces the intensity of Ly$\beta$ rather than that of Ly$\alpha$.
RS by the ejecta may selectively reduce Ly$\alpha$, which we did not take into account in our analysis.
We will be able to evaluate the ejecta contribution by measuring the intensity ratio of the emission lines in Fe K$\alpha$, which originates only from the ejecta \citep[e.g.,][]{Yamaguchi2014, Uchida2015}.
Since Fe K$\alpha$  is out of the wavelength band of the RGS, further observations with X-ray microcalorimeters \citep[e.g.,][]{Kelley2016} is required to clarify this point.

\section{Conclusions} \label{sec:conclusions}
We analyzed a high resolution X-ray grating spectrum of LMC SNR N49 obtained with the RGS aboard XMM-Newton. 
We found that the G-ratio of \ion{O}{7} He$\alpha$ is significantly higher than that expected for a thin thermal plasma emission.
The ratios of \ion{Fe}{17} L$\alpha$ (3s--2p)/(3d--2p) and \ion{O}{8} Ly$\beta$/$\alpha$ also show large residuals from the model.
While an extra CX component well reproduces the G-ratio of the \ion{O}{7} He$\alpha$ triplet, the residuals around \ion{Fe}{17} L$\alpha$ and \ion{O}{8} Ly$\alpha$ 
still remain.
On the other hand, RS can fairly reproduce the RGS spectrum. 
We estimated the optical depth for the RS from the intensities of the scattered lines and found that the depth is roughly consistent with the size of N49.  
Our results indicate that RS has a particularly strong effect on the measurement of oxygen abundance.
As demonstrated by  \cite{Hitomi2018} with the X-ray microcalorimeter SXS aboard Hitomi \citep{Kelley2016}, future missions such as the X-Ray Imaging and Spectroscopy Mission \citep[XRISM;][]{Tashiro2018} and the Advanced Telescope for High ENergy Astrophysics \citep[Athena;][]{Nandra2013} will provide useful means for further studying the effects of RS in SNRs.

\acknowledgments
We would like to thank Dr. Hiroya Yamaguchi for fruitful discussions.
We also thank Dr. John Raymond for helpful advice.
We deeply appreciate all the XMM-Newton team members.
This work is supported by JSPS/MEXT Scientific Research grant Nos. JP25109004 (T.T. and T.G.T.), JP19H01936 (T.T.), JP26800102 (H.U.), JP19K03915 (H.U.), and JP15H02090 (T.G.T.).



\end{document}